\documentclass[longauth]{aa} % for the long lists of affiliations 
\usepackage{subcaption}

\usepackage{graphicx}
\usepackage{xcolor}
\usepackage{hyperref}
\usepackage{txfonts}
\newcommand{\logg}{\ensuremath{\log g}}

\newcommand{\feh}{\rm [Fe/H]}
\newcommand{\teff}{T$_{\rm eff}$}
\def\kms{\,$\mathrm{km\, s^{-1}}$}

\usepackage{color,soul}

\definecolor{lightblue}{rgb}{.70,.95,1}
\sethlcolor{lightblue} 

\begin{document}
\title{The $^{12}$C/$^{13}$C isotopic ratio at the dawn of chemical evolution  \thanks{Based on ESPRESSO GTO observations collected under ESO programmes 1104.C-0350 and 108.2268.001 (P.I. P. Molaro).}
}
\author{P. Molaro\inst{1,2}, %INAF Trieste+Institute Trieste
D.~S. Aguado\inst{3,5}, %INAF Florence
E. Caffau\inst{4}, %OBSPM+Institute Trieste
C. Allende Prieto\inst{3,5}, %IAC+ULL
P. Bonifacio\inst{4}, %OBSPM+Institute Trieste
J.~I. Gonz\'alez Hern\'andez\inst{3,5}, %IAC+ULL
R. Rebolo\inst{3,5,6}, %IAC+ULL+CSIC
%S. Salvadori \inst{1,2},
M. R. Zapatero Osorio\inst{7}, %CSIC-INTA
S. Cristiani\inst{1,2},
F. Pepe\inst{8},
N.~C. Santos\inst{9,10},
Y. Alibert, \inst{16},
G. Cupani\inst{1,2},
P. Di Marcantonio\inst{1}
V. D'Odorico\inst{1,2,11},
C. Lovis\inst{8},
C. J. A. P. Martins\inst{9,13},
D. Milakovi\'{c}\inst{1,2,14},
M. T. Murphy\inst{2,17},
%J. Rodrigues\inst{9,10},
N.~J. Nunes\inst{12},
T.~M. Schmidt\inst{8,1},
S. Sousa\inst{9}
A. Sozzetti\inst{15},
A. Su\'arez Mascare\~no\inst{3,5}
}

%\textit{(Affiliations can be found after the references)}

\institute{INAF-Osservatorio Astronomico di Trieste, Via G.B. Tiepolo 11, I-34143 Trieste, Italy.
\and
Institute of Fundamental Physics of the Universe, Via Beirut 2, I-34151, Trieste, Italy.
\and
Instituto de Astrof\'{\i}sica de Canarias, V\'{\i}a L\'actea, 38205 La Laguna, Tenerife, Spain.
\and
GEPI, Observatoire de Paris, Université PSL, CNRS, 5 Place Jules Janssen, 92190 Meudon, France.
\and
Universidad de La Laguna, Departamento de Astrof\'{\i}sica,  38206 La Laguna, Tenerife, Spain.
\and
Consejo Superior de Investigaciones Cient\'{\i}ficas, 28006 Madrid, Spain.
\and
Centro de Astrobiología (CSIC-INTA), Carretera Ajalvir km 4, 28850 Torrejón de Ardoz, Madrid, Spain.
\and
Département d’astronomie de l’Université de Genève, Chemin Pegasi 51, 1290 Versoix, Switzerland.
\and
Instituto de Astrof\'isica e Ci\^encias do Espa\c co, CAUP, Universidade do Porto, Rua das Estrelas, 4150-762, Porto, Portugal.
\and
Departamento de Física e Astronomia, Faculdade de Ciências, Universidade do Porto, Rua Campo Alegre, 4169-007 Porto, Portugal.     
\and
Scuola Normale Superiore P.zza dei Cavalieri, 7 I-56126 Pisa.
\and
Instituto de Astrofísica e Ciências do Espaço, Faculdade de Ciências da Universidade de Lisboa,
Campo Grande, PT1749-016 Lisboa, Portugal.
\and
Centro de Astrof\'{\i}sica da Universidade do Porto, Rua das Estrelas, 4150-762 Porto, Portugal 
\and
INFN, Sezione di Trieste, Via Valerio 2, I-34127 Trieste, Italy
\and
INAF - Osservatorio Astrofisico di Torino, Via Osservatorio 20, I-10025 Pino Torinese, Italy.
\and
Physics Institute of University of Bern, Gesellschafts strasse 6, CH3012 Bern, Switzerland
\and
Centre for Astrophysics and Supercomputing, Swinburne University of Technology, Hawthorn, Victoria 3122, Australia. 
\\
}   

\authorrunning{Molaro et al.\\}
\titlerunning{ The $^{12}$C/$^{13}$C isotopic ratio at the dawn of chemical evolution}

% \abstract{}{}{}{}{} 
% 5 {} token are mandatory
 
  \abstract
   {The known mega metal-poor (MMP) and  hyper metal-poor (HMP) stars, with [Fe/H] $<$ -6.0 and $<$ -5.0, respectively,    likely belong to the CEMP-no class, namely, carbon-enhanced stars with little or no second peak neutron-capture elements. They are likely second-generation stars, and the few  elements  measurable  in their atmospheres are used to infer  the properties of  a single or very few progenitors. }
  % aims heading (mandatory)of 
   { The high carbon abundance   in the CEMP-no stars offers  a unique opportunity to measure the carbon isotopic ratio, which     directly indicates  the presence  of  mixing between the He- and H-burning layers either within the star or in the  progenitor(s). By means of  high-resolution spectra   acquired with the ESPRESSO spectrograph at the VLT,  we aim to  derive   values  for the $^{12}$C/$^{13}$C   ratio at the lowest metallicities. }
   {We used a  spectral synthesis technique based on the SYNTHE code and on ATLAS models within a Markov chain Monte Carlo  methodology to derive $^{12}$C/$^{13}$C in the stellar atmospheres of  four of the most metal-poor stars known: the   MMP  giant  SMSS\,J0313$-$6708 ([Fe/H] $<$ -7.1),  the HMP dwarf  HE\,1327$-$2326 ([Fe/H] = -5.8), the HMP giant SDSS\,J1313$-$0019 ([Fe/H] = -5.0), and the ultra metal-poor subgiant  HE\,0233$-$0343 ([Fe/H] = -4.7).  We also revised a previous  value for the MMP   giant   SMSS\,J1605$-$1443 ([Fe/H] = -6.2).}
   {In four stars  we derive  an isotopic  value while for HE\,1327$-$2326   we provide  a  lower limit. All measurements are  in the range  39 $<$  $^{12}$C/$^{13}$C$<$ 100, showing that  the He- and H-burning layers underwent  partial mixing either  in the stars or, more likely,    in their  progenitors. This provides evidence of  a primary production of $^{13}$C at the dawn of  chemical evolution.   CEMP-no dwarf stars with slightly higher metallicities show  lower isotopic values, $<$ 30 and even approaching   the CNO cycle   equilibrium value.  Thus, extant data   suggest the presence of  a discontinuity  in the  $^{12}$C/$^{13}$C  ratio  at around [Fe/H]$\approx$  - 4 , which  could mark a real  difference between the progenitor pollution  captured by   stars with different metallicities.  We also note that some  MMP and HMP stars with high $^{12}$C/$^{13}$C  show   low $^7$Li values, providing an indication that mixing in the CEMP-no progenitors is not responsible for the observed Li  depletion.   }
   {}
     
               
\maketitle

\section{Introduction}\label{sec:intro}
Following the discovery of  HD 140283 and HD 19445 by \citet{chamberlain1951ApJ...114...52C}, the search for metal-poor stars  proceeded serendipitously  or by selecting stars with anomalous kinematics  with regard to the Galactic rotation  \citep{bessel1984ApJ...285..622B}. A breakthrough came in the 1980s with the H\&K survey of \citet{beers1985AJ.....90.2089B}, who identified   many  metal-poor candidates with [Fe/H] $\approx$ -4   that were later confirmed by means of  high-resolution follow-ups \citep{molarocastelli1990A&A...228..426M,molarobonifacio1990A&A...236L...5M}. 
%revealing also an
%anomalous spread in the neutron capture elements (Norris et al. 1993; Primas et al. 1994) as
The  H\&K survey  revealed  that a large fraction of metal-poor stars were  enriched in carbon, the so-called carbon-enriched metal-poor (CEMP) stars.   CEMP stars show the presence of neutron-capture (n-capture) elements, revealing the imprint of asymptotic giant branch (AGB) contamination.  
 A puzzling CEMP star without any detectable presence of   elements formed by the slow n-capture process,  CS 22957-27,     was found by \citet{norris1997ApJ...489L.169N}  and \citet{bonifacio1998A&A...332..672B}. These papers  were submitted  a few days   apart.   
 %The prototype of the CEMP-no is  CS 22957-027   discovered   almost simultaneously by \citet{norris1997ApJ...489L.169N} and \citet{bonifacio1998A&A...332..672B}. 
 %This star
 % has also been shown to be a long period ($\rm P=3125$\,d) non interacting binary system  \citet{preston2001AJ....122.1545P}. 
  Other  CEMP stars  without signs of n-capture elements      were found by \citet{aoki2002ApJ...567.1166A},   and
\citet{chris04} discovered  the remarkable  HE\,0107$-$5240 giant with  $\rm [Fe/H]=-5.4$, [C/Fe] $>$ 3.9, and no detectable  n-capture elements,  from the  Hamburg/ESO objective prism survey.   These stars   were grouped into  the CEMP-no class,   where the  `no' indicates the absence of n-capture elements,   defined as stars  with [C/Fe]$>$ 0.7, originally $>$ 1.0,   and [Ba/Fe]$<$ 0.0;       the more common  stars     with  [Ba/Fe] $>$ 1  are called CEMP-s, where `s' indicates the presence of n-capture elements \citep{beers2005ARA&A..43..531B}.

\citet{spi13} noted that  [C/Fe]  increases with  decreasing  iron, which implies a constant   carbon abundance, and  suggested  two  different nucleosynthetic origins for the carbon in the CEMP-no and CEMP-s stars. 
%In the latter   there is a  high C level, almost  solar,  while in the CEMP-no    C  is  $\approx$  -1.5 dex below solar. The two groups are more clearly separated at low metallicities while  they merge at about [Fe/H] $\approx$ -3 due to the general increase of C.
While in  CEMP-s  stars carbon is coming from
an AGB companion along with the  n-capture elements, 
 in the CEMP-no  stars the carbon was  presumably already present in the gas from which the star formed \citep{spi13, boni15}.  Radial velocity studies  have found that
 almost all  CEMP-s stars  are  in binary systems, unlike CEMP-no stars, which supports this idea \citep{hansen2015ApJ...807..173H, han16I}. 
 %The few exceptions found could
%be  explained if few systems are seen
%pole-one or if there are binaries with
%stars which do not evolve into an AGB phase. 

Subsequent surveys such as  the Sloan Digital Sky Survey \citep{york2000AJ....120.1579Y} and its extensions SEGUE \citep{yanny2009AJ....137.4377Y} and SEGUE-2 \citep{rockosi2022ApJS..259...60R}, AEGIS \citep{kell07}, LAMOST \citep{deng2012RAA....12..735D},  Pristine \citep{starkenburg2017MNRAS.471.2587S}, and TOPoS \citep{caffau2013A&A...560A..71C} increased the number of known CEMP-no stars, showing that they  dominate the low-metallicity tail. Metal-poor stars are defined as ultra metal-poor (UMP), hyper metal-poor (HMP), and mega metal-poor  (MMP),  corresponding to $\rm [Fe/H]<-4$,
$\rm[Fe/H]<-5$, and $\rm[Fe/H]<-6$, respectively. There are    two MMP stars and  seven  HMP stars known to date,  all of which are CEMP-no.  There are also five known UMP stars   with [Fe/H] $<$ -4.5,
%\citep{caff16,han15I,starkenburg14,norris2013ApJ...762...28N}
three of which  are  CEMP-no and two are  normal stars with approximately solar-scaled  abundances (see \citealt{Aguado2023A&A...669L...4A} and references therein). 
%\citep{chris04,fre06,kel14,caff16,boni15,boni18,agu19a,nordla%nder2019MNRAS.488L.109N,alle15}. 
Cooling of the gas   by fine-structure line emission  of singly ionised carbon or neutral atomic oxygen  could have allowed the formation of CEMP-no stars   earlier than  those      with normal carbon levels \citep{volker2003Natur.425..812B}.  %However,
%the very low metallicity stars with normal carbon   call for a more complex mechanism such as dust cooling \citep{sche12} or turbulent fragmentation \citep{gre12}.  
CEMP-s stars are seen only for [Fe/H] $>$ -4.5, and this is likely  because CEMP-s stars require the time necessary for  the companion to evolve into an  AGB star.
It was also determined that  CEMP-no stars are found primarily in the outer halo \citep{car12}, while a significant dearth of CEMP-no stars in the Galactic Bulge is interpreted as a  signature of pair instability supernovae \citep[SNe;][]{pagnini2023MNRAS.521.5699P}. \citet{zepeda2022} found evidence that  CEMP-no stars form chemo-dynamically  tagged groups with a low dispersion in their [C/Fe] abundances.

The extremely low iron abundance of MMP and HMP stars
suggests that only a few and perhaps just one single   progenitor
polluted the gas out of which they 
formed. The C enhancement could originate  from
faint SNe with energies of 10$^{51}$ erg together
with mixing and fallback  \citep{ume03,umeda2005,Nomoto2013,tominaga2007,tominaga2014ApJ...785...98T}. With a  large amount of fallback, faint SNe eject a lot of carbon and a small amount of Fe. This   produces 
  ejecta with large [C/Fe] abundance ratios. An alternative model  to explain the CEMP-no stars is represented by   massive, low-metallicity fast-rotating stars called spinstars, with efficient mixing and mass loss \citep{Meynet2006A,chiappini2008A&A...479L...9C,maeder2015A&A...576A..56M,Limongi2018}.
Therefore, the
identification of the progenitors of CEMP-no stars has important bearings on the nature of the first stars \citep{fre15rev}. The elemental abundance pattern  includes a limited number  of elements, and 
the progenitors are poorly constrained. Thanks to their carbon enhancement,  CEMP-no stars offer a   unique opportunity to explore the behaviour of  $^{12}$C/$^{13}$C  at the lowest metallicities.    
At low metallicities, very high $^{12}$C/$^{13}$C ratios, greater than 10$^3$,  are expected due to the secondary nature of $^{13}$C \citep{chiappini2008A&A...479L...9C,Romano2003MNRAS.342..185R,romano2022A&ARv..30....7R,kobayaschi2020ApJ...900..179K}.  \citet{aguado22,Aguado2023A&A...669L...4A} derived  significant   bounds to  the isotopic ratio $^{12}$C/$^{13}$C in HE\,0107$-$5240  and  SMSS\,J1605$-$1443, with    $\feh=-5.8$ and $-6.2$, respectively.
 In this paper we extend this  analysis to   four   additional known extremely  metal-poor (EMP) CEMP-no stars. We also  reconsider the previous analysis of SMSS\,J1605$-$1443,  for which new observations have since been conducted,  and turn the bound into  a value. We then discuss the results in the context of $^{12}$C/$^{13}$C determinations  in   other CEMP-no stars, highlighting some  implications  that had until now been overlooked.

\section{Observations and data reduction}\label{sec:observations}
Observations of  some of the most metal-poor stars known have been  taken   with the  ESPRESSO, the Echelle SPectrograph for Rocky Exoplanets and Stable Spectroscopic  Observations,  at the Very Large Telescope within the  ESO Guaranteed Time Observations (GTO) programme on EMP stars (PI Paolo Molaro). The stars are    SMSS\,J0313$-$6708,    HE\,1327-2326, HE\,1313-0019  (also known as       2MASS J13132688-0019415), and  
 HE\,0233-0343 (also known as SDSS J002314.00+030758.0). Since a few additional observations for  SMSS\,J1605$-$1443  were taken after the \citet{Aguado2023A&A...669L...4A} analysis, we also revised the carbon isotopic ratio estimation for this star. The journal of the observations is given in Table \ref{table1} together with the ESPRESSO observing mode and the exposure times.
The ESPRESSO spectrograph has  two fibres, one for the target and one for the  sky,  with diameter of $140\,\mu m$ that corresponds to a 1\farcs0 aperture on sky   and  provides a spectral coverage from  380.0   to  780.0  nm \citep{pepe21}. The CCD was binned by either $4\times2$ pixels (i.e. 4 pixels  binned in the spatial  direction and 2  in the spectral one)  or by $2\times1$ pixels, as reported in the third column of Table  \ref{table1}. The  corresponding  resolution was R = $\lambda$ / $\delta \lambda$ $\approx$ 140.000 for both binnings. The observations   were performed in service mode with  individual exposure times of the order of $3000$\,s. The precise exposure times and the signal-to-noise ratio per pixel measured at 430,0 nm are provided in the last two columns of   Table \ref{table1}.

The automatic Data Reduction Software, DRS  3.0, of the ESPRESSO pipeline was used for the data reduction, which includes  bias,  flat-fielding correction, and sky subtraction, with the sky  taken from the second fibre. The wavelength calibration combines a ThAr lamp  with a Fabry-P\'erot etalon  as described in \citet[][]{pepe13}.

%-----------------------------------------------------

%--
%-----------------------------------------------------------------

\begin{table}[t!]  %tbl
\vspace*{0mm}
%\hspace*{0cm}\resizebox{1.\linewidth}{!}{
 \caption{Journal of observations.  } 
 \label{table1}
 \begin{center}\scriptsize
\begin{tabular}{lccccc}
 \hline
 STAR  & MJD\tablefootmark{a} & ESPRESSO MODE&t$_{exp}$&S/N&\\
    & & &s &\@4300\,\AA \\
 \hline
  \hline
SMSS\,J0313$-$6708  & 58701.281 & HR21 &3400&12      & \\
SMSS\,J0313$-$6708  & 58732.245 & HR21 &3400&10      & \\
SMSS\,J0313$-$6708 & 58732.343 & HR21 &3400&14      & \\
SMSS\,J0313$-$6708  & 58740.153 & HR21 &3400&9      & \\
SMSS\,J0313$-$6708  & 58740.196 & HR21 &3400&8      & \\
SMSS\,J0313$-$6708  & 58742.119 & HR21 &3400&7      & \\
SMSS\,J0313$-$6708  & 58743.268 & HR21 &3400&15      & \\
SMSS\,J0313$-$6708  & 58743.335 & HR21 &3400&12      & \\
SMSS\,J0313$-$6708  & 58761.310 & HR21 &3400&14      & \\
%SMSS\,J0313$-$6708 & 58780.153 & HR21 &3417&--      & \\
SMSS\,J0313$-$6708  & 58793.309 & HR21 &3400&16      & \\
SMSS\,J0313$-$6708 & 59190.052 & HR42 &3417&20      & \\
SMSS\,J0313$-$6708 & 59237.078 & HR42 &3417&13      & \\
SMSS\,J0313$-$6708  & 59250.041 & HR42 &3417&11      & \\
SMSS\,J0313$-$6708  & 59264.017 & HR42 &3000&21      & \\
%SMSS\,J0313$-$6708 &59499.118&HR42&3000&--\\
SMSS\,J0313$-$6708 &59513.248&HR42&3000&23\\
SMSS\,J0313$-$6708 &59882.103&HR42&3144&11\\
SMSS\,J0313$-$6708 &59884.124&HR42& 3144&17\\

HE\,1327$-$2326    & 58615.137 & HR21 &1800&20      &\\
%HE\,1327$-$2326   & 63.287   &  0.106 & 8667.160 & HR21 &1800&      & Low SNR 4!\\
HE\,1327$-$2326   & 58688.022 & HR21 &2700&23      &\\
HE\,1327$-$2326    & 58688.055 & HR21 &2700&22      &\\
HE\,1327$-$2326   & 58695.996 & HR21 &1800&25      &\\
HE\,1327$-$2326   & 58697.988 & HR21 &2400&18      &\\
HE\,1327$-$2326   & 58841.331 & HR21 &1800&23      &\\
HE\,1327$-$2326   & 59429.054 & HR42 &3000&16      &\\
HE\,1327$-$2326  & 59663.072 & HR42 &2844&41      &\\
HE\,1327$-$2326  &59990.288&HR42&3144& 42   \\
HE\,0233$-$0343  & 58699.288 & HR21 &3400&10      & \\
HE\,0233$-$0343   & 58760.318 & HR21 &3400&11      & \\
HE\,0233$-$0343   & 58783.112 & HR21 &3400&14      & \\
HE\,0233$-$0343   & 58813.243 & HR21 &3400& 14     & \\
SDSS\,J1313$-$0019 & 59649.249 & HR42 &3444&12      & \\
SDSS\,J1313$-$0019 & 59972.310 & HR42 &3144&13      & \\
SDSS\,J1313$-$0019 & 60077.193 & HR42 &3144&9      & \\
SDSS\,J1313$-$0019 & 60080.092 & HR42 &3144&11      & \\
SDSS\,J1313$-$0019 & 60105.121 & HR42 &3144&12      & \\
SMSS\,J1605$-$1443  & 59676.277 & HR42&3150&11\\    
SMSS\,J1605$-$1443  & 59726.189 & HR42&3150&11 \\  
SMSS\,J1605$-$1443 & 59727.060 & HR42&3150&14\\    
SMSS\,J1605$-$1443 & 59761.059 & HR42&3150&10\\    
SMSS\,J1605$-$1443 & 59792.055 & HR42&3150&15\\  
SMSS\,J1605$-$1443 & 60030.365 & HR42&3144&10\\ 
SMSS\,J1605$-$1443 & 60078.265 & HR42&3144&14\\ 

\hline
\end{tabular}
\tablefoot{\tablefoottext{a}{Modified Julian date at the start of the observations.}}
\end{center}
\end{table}

\begin{figure}
\begin{center}
{\includegraphics[width=90 mm, angle=00,trim={ .cm .cm .3cm 0cm},clip]{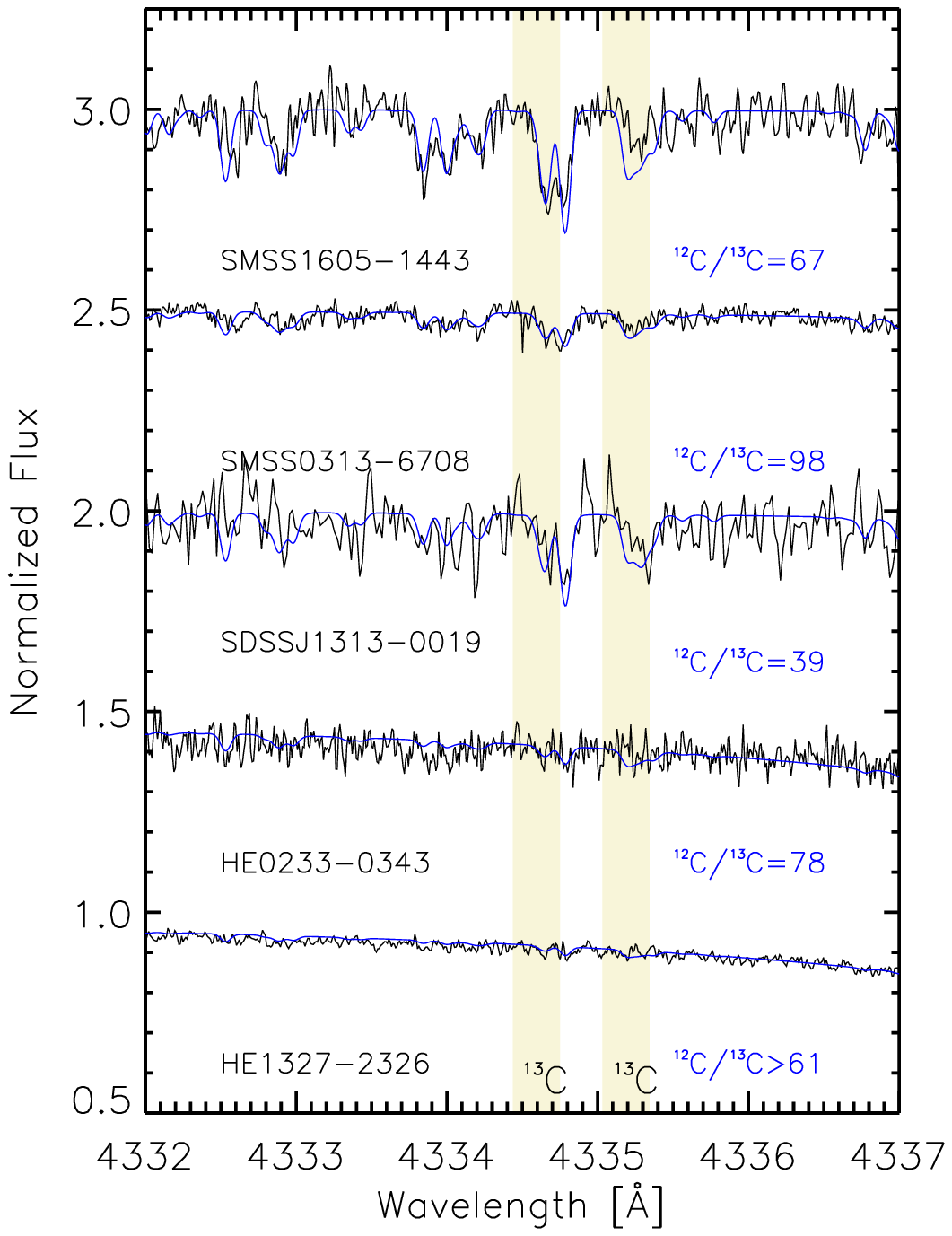}}
%{\includegraphics[width=65 mm, angle=90]{ /mcmc_v2.eps}}
\end{center}
\caption{  Portion of the combined ESPRESSO spectra of the HMP stars, with a focus on the strongest lines of  $^{13}$C in the G band. The best fit derived with FERRE using an MCMC algorithm is displayed in magenta. 
%Lower-panel: Distribution of the Markov Chain Monte Carlo   experiments versus the most likely  $^{12}$C/$^{13}$C value.
}
\label{fig:fig1_spectrum}
\end{figure}

\section{Analysis}\label{sec:analysis}
\subsection{The programme stars}

  In order to study the $^{12}$C/$^{13}$C isotopic ratio at the beginning of chemical evolution, we selected some of the most metal-poor stars known. The sample includes the giant SMSS\,J0313$-$6708, discovered by \citet{kel14}; with  $\rm [Fe/H]<-7.3$, it is the most iron-poor star presently known. We note that the inferred iron abundances  are significantly higher  in 3D non-local thermodynamic equilibrium than in 1D thermodynamic equilibrium, by 0.8 dex \citep{nor17}. The second most metal-poor giant  is SMSS\,J1605$-$1443,  first detected by the SkyMapper telescope; its     iron abundance  is measured at $\feh=-6.2$ by \citet{nordlander2019MNRAS.488L.109N}. HE\,1327$-$2326 with  $\rm [Fe/H]=-5.60$  is the most metal-poor  turn-off star. It was   discovered  by \citet{fre05,fre08} and studied by  \citet{aoki06I}. The star has  [Zn/Fe] = 0.8, and  a 25 M$\sun$  aspherical SN model  exploding with E = 5 $\times$  10$^{51}$ erg has been suggested  to provide the best match with the high zinc abundance  \citep{ezzeddine20}.
 Our programme  sample  also includes the giant SDSS\,J1313$-$0019 with a  measured metallicity of $\rm [Fe/H]=-5.0 \pm 0.1$. This star was   discovered by \citet{alle15} and  studied by \citet{fre15} and \citet{agu17II}.   \citet{alle15} and \citet{fre15} suggested that SDSS\,J1313$-$0019 could be a binary system. Finally, the sample includes another     turn-off star, HE\,0233$-$0343,    with  \rm [Fe/H]=-4.68 $\pm$ 0.2 dex \citep{beers2007ApJS..168..128B,garcia2008PhST..133a4036G,han14}. 
The carbon abundance  in all these stars is $\rm A(C)\approx 6-7$, falling in the low-C band; although their barium  abundance limits are not sensitive enough to match the condition [Ba/Fe]$<$ 0, all of them can be classified  as  CEMP-no stars.

\begin{figure}
\begin{center}
{\includegraphics[width=65 mm, angle=90]{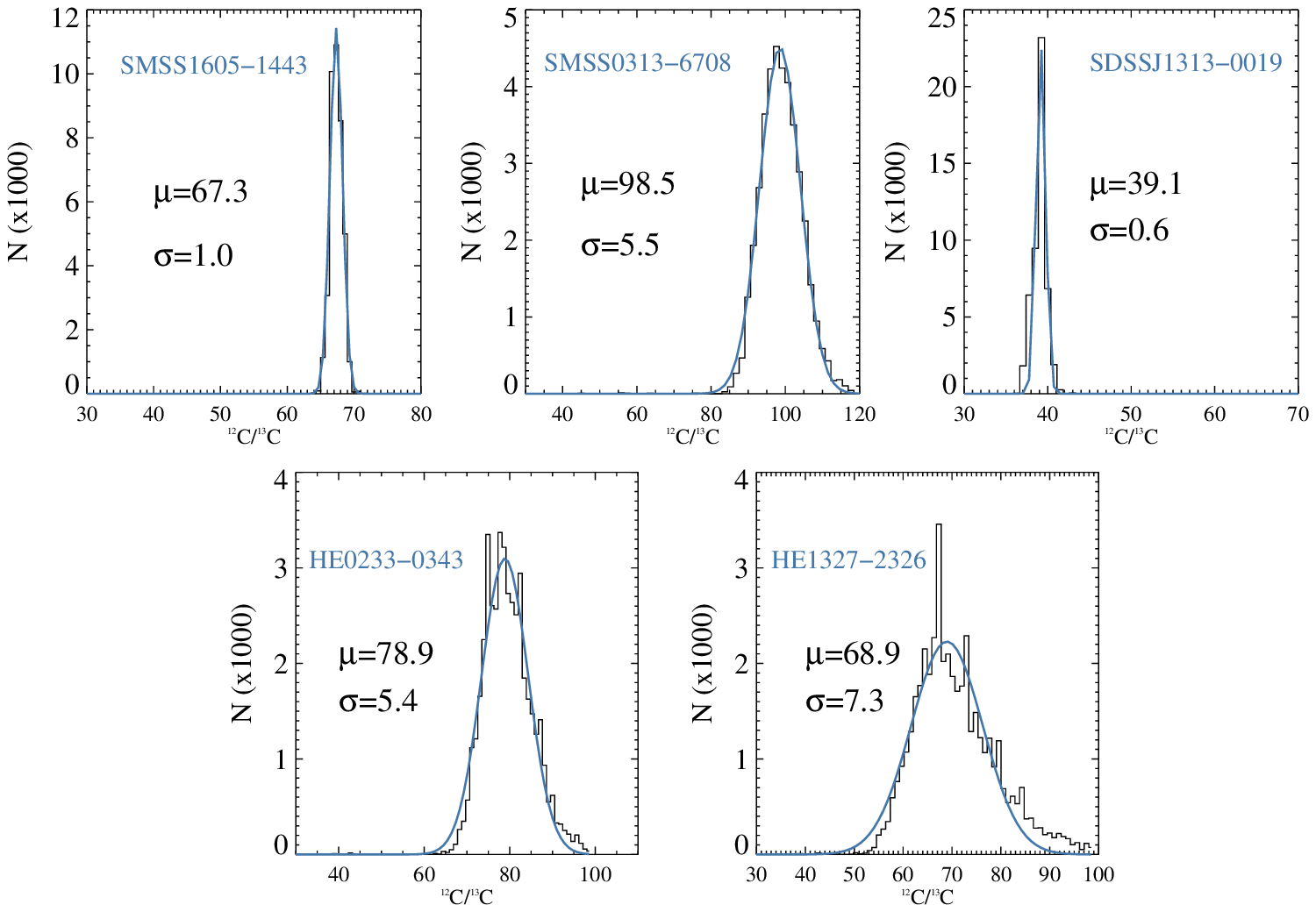}}
\end{center}
\caption{ 
Distribution of the MCMC  experiments versus the most likely $^{12}$C/$^{13}$C value for each of the  targets. The mean value and typical deviation are displayed to provide a summary of the results.}
\label{fig:fig2}
\end{figure}

\subsection{Carbon isotopic ratio $^{12}$C/$^{13}$C}\label{caff}

Deriving $^{12}$C/$^{13}$C ratios in metal-poor stars is  challenging due to the weakness of the $^{13}$C lines. However,  the high carbon abundance of  CEMP stars and  the hundreds of lines of the G band  collectively provide sufficient  information \citep{masseron2014A&A...571A..47M}.  Portions of the combined ESPRESSO spectra of the programme stars corresponding to the strongest $^{13}$C lines are shown in Fig. \ref{fig:fig1_spectrum}.  We computed stellar models using the {\tt ATLAS} model atmosphere code  assuming stellar parameters such as effective temperature (\teff), surface gravity (\logg), metallicity (\feh), and total carbon abundance, A(C). This code employs an opacity distribution function corresponding to a metallicity of $-5.0$
\footnote{\url{https://wwwuser.oats.inaf.it/castelli/odfnew.html}}.

The stellar parameters used in our analysis are those obtained from the most complete studies of the stars under study since there is no way to verify or improve them.
The adopted stellar parameters for SMSS\,J0313$-$6708 are \teff = 5125 $\pm$ 100 K and  $\log g$ = 2.3 $\pm$ 0.2,    as derived from low-resolution spectrophotometry by \citet{kel14}. Those of 
  HE\,1327$-$2326   are   \teff = 6180 $\pm$ 100 K and  $\log$ g = 3.70 $\pm$ 0.3, from \citet{aoki06I}.  We note that \citet{aoki06I} also find a solution for a slightly higher surface gravity of $\log g$ =4.5 $\pm$ 0.3  with an almost equal probability that the star is  a dwarf or a subgiant, though the \textit{Gaia} parallax favours the subgiant solution.  
 For HE\,0233$-$0343,   \citet{hansen2015ApJ...807..173H} obtained   \teff = 6100 $\pm$100 K and  $\log g$ = 3.4 $\pm$ 0.3 dex, but the \textit{Gaia} parallax  suggests  T$_{eff} = 6230$  and  $\log g $ = 4.43.
SDSS\,J1313$-$0019    has   T$_{eff} = 5200 \pm 150$  K and $\log g = 26 \pm 0.5 \pm$  according to \citet{fre15}.  
   The stellar parameters of  SMSS\,J1605$-$1443   are  \teff =4850 $\pm$ 100  and  $\log g$ = 2.0 $\pm$ 0.2 dex as derived by \citet{nordlander2019MNRAS.488L.109N}. These stellar parameters are summarised in Table \ref{table2}.
We adopted an [$\alpha$/Fe] ratio of +0.4 and microturbulences of 1.8, 2.0, 1.5, 1.5, and 1.5\,km$\,$s$^{-1}$ for SMSS\,J1605$-$1443, SMSS\,J0313$-$6708, SDSS\,J1313$-$0019, HE\,0233$-$0343, and HE\,1327$-$2326, respectively. 
It should be noted that line intensities of isotopic molecules originate from identical or very similar rotational and vibrational levels, making the isotope ratio relatively insensitive to uncertainties in the atmospheric parameters. 

With this model atmosphere,  we generated a grid of synthetic spectra using the  SYNTHE code \citep{kur05, sbo05}. We adopted the\ solar carbon abundance of A(C)$_{\odot}=8.46$ from \citet{asplund2021A&A...653A.141A}. For each set of stellar parameters corresponding to our targets, we computed models with varying $^{12}$C/$^{13}$C ratios: 2.2, 5.3, 11.6, 24.1, 49.1, 99.0, and 198.6. These ratios were specifically chosen to capture the largely non-linear behaviour of  $^{12}$C/$^{13}$C ratios.
With the SYNTHE stellar models, a grid in a format compatible with the  FERRE code   was constructed \citep{alle06} \footnote{available at \url{http://github.com/callendeprieto/ferre}}. We generated one grid per target, allowing the FERRE code to read and process the data seamlessly. 

In  \citet{Aguado2023A&A...669L...4A}  we tested a Markov chain Monte Carlo (MCMC) technique that enables the derivation of $^{12}$C/$^{13}$C ratios even with marginal detections of the $^{13}$C lines. This methodology is described in Appendix B of \citet{Aguado2023A&A...669L...4A}. In short, we identified the spectral regions of  the G band near 430.0 nm, where the $^{13}$C information is maximum, by  subtracting two theoretical models with high and low $^{12}$C/$^{13}$C, respectively. This enabled us to concentrate the fitting procedure of the code exclusively on the pertinent regions while disregarding other features or minor artefacts present in the data. After  a 300-pixel running-mean normalisation, the code employs an MCMC self-adaptive randomised subspace sampling algorithm for the fitting, which is  described in \citet{vru09}. Ten chains of 5,000 experiments each were run following this MCMC methodology. Based on our experience with the code, experiments from the initial chain can encounter a blockage. To address this, we allowed the algorithm to ignore the first 500 iterations. Finally, the code provides us with the most probable result and the samples from the Markov chains, following the approach used in \citet{Aguado2023A&A...669L...4A}. The sample distributions of the MCMC experiments versus the most likely $^{12}$C/$^{13}$C value are shown in Fig. \ref{fig:fig2}.  
In Fig. \ref{fig:fig1_spectrum} the best FERRE fits are compared with the data in the region around $\sim422.5$\,nm  at the edge of the G band, where the strongest $^{13}$C absorptions are.  

 We successfully measured the $^{12}$C/$^{13}$C ratios  in the three relatively cool giants. For the two MMP stars -- SMSS\,J0313$-$6708 and HE\,1605-1443 -- the $^{12}$C/$^{13}$C ratios are 98.5 $\pm$ 5.5 and 67.3 $\pm$ 1.0, respectively. Previously, \citet{Aguado2023A&A...669L...4A} had set a lower limit of $^{12}$C/$^{13}$C $>$ 60 for HE 1605-1443. In the HMP SDSS\,J1313$-$0019, the $^{12}$C/$^{13}$C ratio is 39.1 $\pm$ 0.6. For the two relatively warm and unevolved stars, HE\,1327-2326 and HE\,0233$-$0343, the code provides $^{12}$C/$^{13}$C ratios of 68.9 $\pm$ 7.3 and 78.9 $\pm$ 5.4, respectively. However, in the spectrum of HE\,1327-2326,  only very faint evidence of the $^{13}$C lines can be observed by eye in Fig. \ref{fig:fig1_spectrum}. Thus, we conservatively considered a lower limit of 47 for this star at the 3$\sigma$ confidence level (CL).
Overall, all the CEMP-no stars investigated in this study exhibit $^{12}$C/$^{13}$C values or lower limits in the range 39 $<$ $^{12}$C/$^{13}$C $<$ 100, providing evidence of primary production of $^{13}$C at the dawn of chemical evolution.

\begin{figure}
\begin{center}
{\includegraphics[width=65 mm, angle=90]
{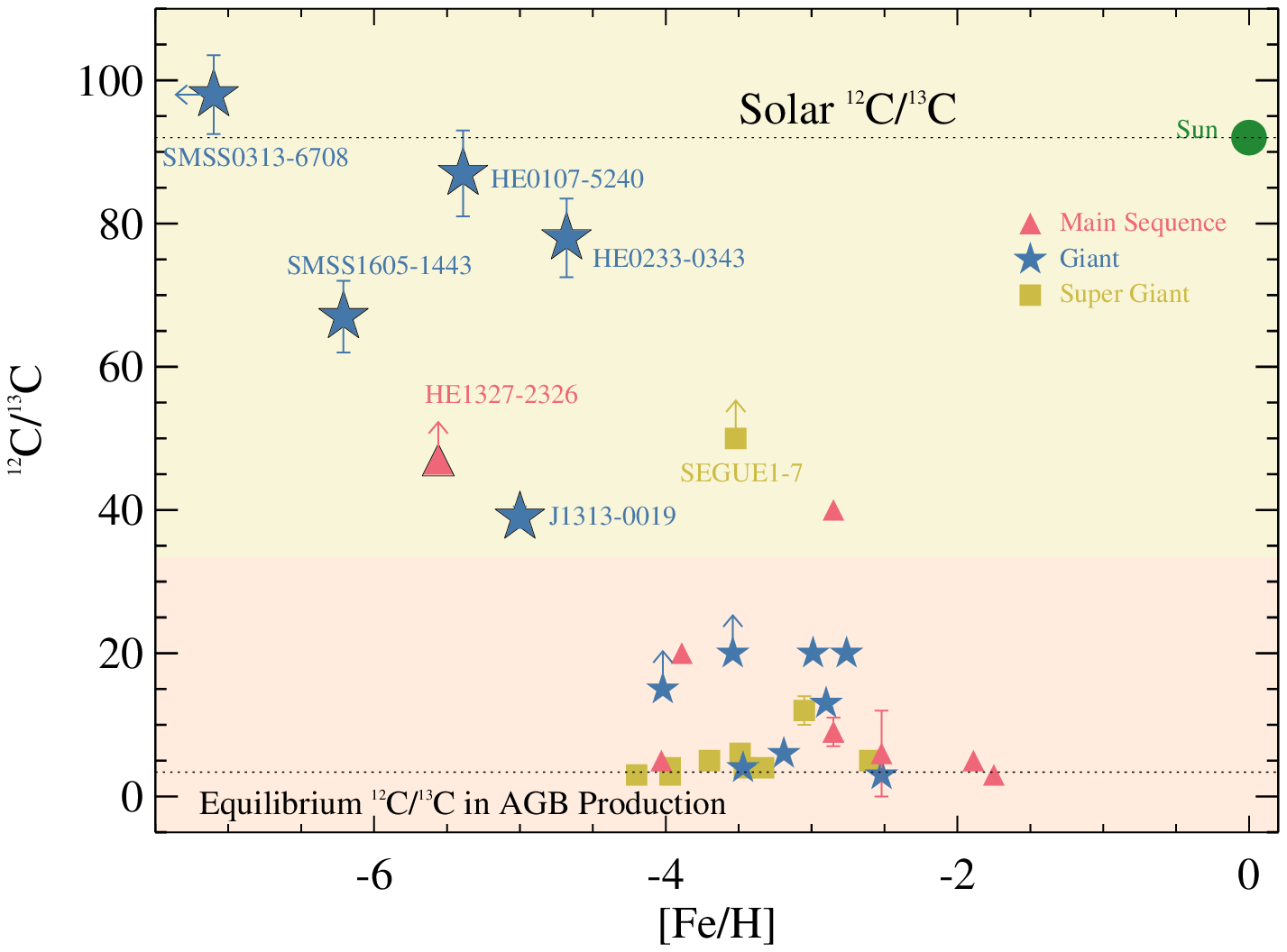}}
\end{center}
\caption{ $^{12}$C/$^{13}$C$-$[Fe/H] plane  for CEMP-no stars. The value for HE\,0107$-$5240 is taken from \citet{aguado22}. Only lower limits greater than 15 are shown since the smaller ones do not provide useful information. Errors in [Fe/H] are omitted for clarity. Red triangles  are  dwarfs with \logg $>$ 3.5. Blue   star-symbols  are stars on the first ascent to the       red giant branch, with  2 $<$ \logg $<$ 3.5. Orange squares  are supergiants with \logg $<$ 2.  The $^{12}$C/$^{13}$C$=30$ value that splits stars with evidence of internal mixing from the others according to \citet{spite06} is also shown; the background above this value is yellow and below is peach coloured. }
\label{fig:iso}
\end{figure}

\section{Discussion}\label{sec:discussion}

The $^{12}$C/$^{13}$C ratio is  a solid indicator
of the presence of CNO-cycle-processed material at
the surface of a star. This processed material could come from the interior of the star or from the progenitors that polluted the gas from which the star formed.
$^{12}$C  is formed in the triple-$\alpha$ process during hydrostatic helium burning and is a primary product of stellar nucleosynthesis. The stable$^{13}$C isotope is  produced in the hydrogen-burning shell when the CN cycle converts pre-existing $^{12}$C into $^{13}$C via proton capture followed by $\beta$ decay.  The $^{12}$C/$^{13}$C  is not locked to the C abundance and provides independent information on either the  evolution of the star or the nature of the progenitor.

As a  star evolves off the main sequence,  the outer convective envelope expands
inwards  into the CN-cycle-processed regions. This mixing episode, called the `first dredge-up',   lowers the $^{12}$C/$^{13}$C ratio from  the   original value \citep{iben1984PhR...105..329I}.

Mixing  also occurs  in the thermal pulses of intermediate-mass stars that become  AGB stars.   This could    lead to a CN-cycle equilibrium ratio of about  $^{12}$C/$^{13}$C  $\sim$3.4, and in fact ratios as small as 6 are observed in the atmospheres of red  supergiants \citep{lambert1977ApJ...215..597L,dearborn1975ApJ...200..675D}. 
 
Nova explosions at temperatures $>$ 10$^8$ K can produce $^{13}$C in a  hot CNO cycle. Pre-nova white dwarfs are rich in CNO nuclei, and the hot CNO cycle triggers the explosion and transmutes CNO into $^{13}$C, $^{15}$N, and $^{17}$O with enhancement factors of 100 times the solar values \citep{starrfield1972ApJ...176..169S, jose1998ApJ...494..680J}. 
All stars as small as 1 solar mass can thus release matter enriched in $^{13}$C into the interstellar medium via stellar wind or at the end of their evolution.\ Therefore,   chemical evolution models  predict a monotonic decrease in the isotopic ratio $^{12}$C/$^{13}$C with time (\citet{Romano2003MNRAS.342..185R,kobayashi20b}).  In fact,   the   value found in the interstellar medium and in  young molecular clouds is  $^{12}$C/$^{13}$C$\sim 60-70, $ which is lower than  the  solar ratio,  $^{12}$C/$^{13}$C$ = 91\pm 1.3$ \citep{goto03,ayres2013ApJ...765...46A}. Moreover,  relatively low values of about 24 have been derived in the Galactic centre by \citet{halfen2017ApJ...845..158H}, and a gradient  is observed as a function of galactocentric distances that clearly shows the secondary nature of $^{13}$C \citep{yan2023A&A...670A..98Y}.

At low metallicities or before the birth of the Solar System, the  $^{12}$C/$^{13}$C  is predicted to be  higher than  solar \citep{Romano2003MNRAS.342..185R,chiappini2008A&A...479L...9C,kobayaschi2020ApJ...900..179K,romano2022A&ARv..30....7R}.
This prediction  is difficult to verify  because there are few observations  that can measure  ratios higher than, or even close to, solar. \citet{botelho2020MNRAS.499.2196B}  in 63 solar twins found  a  value close to solar with  a mild increase in the   metallicity range -0.2$<$ [Fe/H]$<$ -0.0,  which is opposite to what is expected. By contrast,   \citet{crossfield2019ApJ...871L...3C} measured $^{12}$C/$^{13}$C = 296${\pm 45}$ and 224 ${\pm 26}$ in the two components of the  brown dwarf system GJ 745 at [Fe/H] =-0.48, which are too high for the Galactic chemical evolution models.  At lower metallicities,  the isotopic  ratio in   metal-poor giants is found in the range   30-50  and  in the  metal-poor dwarf  HD 140283   is     $^{12}$C/$^{13}$C = 33$^{+12}_{-6}$ \citep{spite06,spite2021A&A...652A..97S}.
%and argued for an  enrichment of 13C in the early Galaxy by massive AGB or fast massive rotators. 
Overall,   observations   do not provide a comprehensive  evolutionary curve of the $^{12}$C/$^{13}$C  in the Galaxy.

In Fig. \ref{fig:iso} we compile  the   existing $^{12}$C/$^{13}$C measurements   or significant  limits  of CEMP-no   stars from the literature.  Data are from \citet{bonifacio1998A&A...332..672B}, \citet{aoki2002ApJ...567.1166A}, \citet{Masseron2012}, \citet{norris2013ApJ...762...28N}, \citet{hansen2015ApJ...807..173H}, and \citet{aguado22,Aguado2023A&A...669L...4A}.  Only lower limits in $^{12}$C/$^{13}$C greater than 15,  which help in differentiating between low and high isotopes ratios, are shown.
 The majority  of the CEMP-no stars  in Fig. \ref{fig:iso} have low $^{12}$C/$^{13}$C values. The CEMP-no supergiants    have $^{12}$C/$^{13}$C $<$ 10, and   the second dredge-up could have  internally synthesised $^{13}$C.  However, quite remarkably there are  low values in four dwarfs that   have not gone through   the first or second dredge-up. They are CS 22958-042 with $^{12}$C/$^{13}$C $=$ 4.5   measured by \citet{sivarani2006}, CS 22887-048 with $^{12}$C/$^{13}$C $=$ 3.7 measured by \citet{aoki2002ApJ...567.1166A}, CS 22956-028 with $^{12}$C/$^{13}$C $=$ 4.3  measured by \citet{masseron2010A&A...509A..93M},  and G77-61 with $^{12}$C/$^{13}$C $=$ 5 measured by \citet{plez2005A&A...434.1117P}.  
 A chemical transfer from a possible massive companion can be ruled out since   all these stars are single or  belong to non-interacting binaries, and,    unless they are  wrongly classified as CEMP-no stars,  the enhancement of  $^{13}$C needs to come from their progenitors. This is probably also true for some giants that show very low  $^{12}$C/$^{13}$C  values that cannot be the result of the first dredge-up.
 %In disk normal main sequence stars $^{12}$C/$^{13}$C values are greater than 50 and are close to the initial isotopic ratio \citep{dearborn1975ApJ...200..675D}.  
 Already in the CEMP-no prototype CS 22957-027,   the $^{12}$C/$^{13}$C ratio is 10 $\pm$ 5, and    \citet{bonifacio1998A&A...332..672B} pointed out that the  star's low luminosity means that the $^{13}$C  could not come from  self-polluting dredge-up processes  and   that the absence of n-capture elements is evidence against mass transfer from an AGB  companion.
  
We note that  these measurements point to a complex Galactic behaviour for the ratio $^{12}$C/$^{13}$C. If it is very   low at metallicities  -3.5 $<$[Fe/H]$<$ -2,  then it  needs to rise  to match the solar value. This implies that the sources of primary $^{13}$C should cease their contribution around [Fe/H] $\approx$ -2.  If so,  when the $^{12}$C  abundance increases  by two orders of magnitude,   the  $^{12}$C/$^{13}$C ratio would also increase by the required  two orders of magnitude.  Unfortunately, the lack of $^{12}$C/$^{13}$C ratio  measurements  in the range -2.0 $<$ [Fe/H] $<$ 0  prevents us from verifying this prediction.  These measurements would have to be extended to other CEMP-no main-sequence stars to see if the low isotope ratios are  common at low metallicities  and   trace the rise of the isotope ratio until it reaches the solar value. 

 Remarkably,  all the  MMP and HMP CEMP-no stars have  $^{12}$C/$^{13}$C $>$ 40,   while  the majority of other CEMP-no  have values $<$ 30. There is also a hint of a smooth decrease in the values going from the MMP to the HMP stars and then to the UMP stars.   In the sample of UMP and EMP stars,  there are   30 CEMP-no stars and all but two   of them show   low $^{12}$C/$^{13}$C ratios, in a few cases close to the CNO equilibrium value.  The exceptions  are  Seque1-7 and   CS31080-095, which have values of   $^{12}$C/$^{13}$C $>$ 50 and  40, respectively,     which is comparable to the more metal-poor  stars.

The extremely low iron abundance of MMP and HMP stars suggests that only a few and perhaps only a single   progenitor polluted the gas out of which they  formed.  The carbon   might originate  from faint SNe with energies of 10$^{51}$ erg  \citep{ume03}. With mixing and a  large amount of fallback, faint SNe eject a lot of carbon and a small amount of Fe, thus   producing  ejecta with large [C/Fe] abundance ratios.  Supernovae with stellar masses of 10-20\,M$_{\odot}$ and low explosion energies of about 10$^{51}$\,erg  have been shown  to be able to reproduce  the pattern of chemical abundances of  CEMP-no stars  \citep{tominaga2014ApJ...785...98T,Almusleh2021AN....342..625A}.   \citet{tominaga2007} successfully reproduced the observed abundance pattern of the CEMP-no star CS 29498-043 \citep{Aoki2004} with a 25 M$_{\odot}$ faint SN model.  \citet{Almusleh2021AN....342..625A}  suggest that  weak SN  progenitors  with stellar masses of 11-22 M$_{\odot}$ and explosion energies in the range 0.3-1.8$\times$10$^{51}$ erg could be the progenitors of the chemical abundances of five CEMP-no stars.  However,  models are not unique, and, for instance, several different kinds of progenitors with masses in the range 12 to 60 M$\sun$ \citep{ishi14,nor17} have been proposed to explain the abundances of  the most iron-poor  star known, SMSS\,J0313$-$6708.
Moreover,  a large dispersion in the [Fe/Ca] values is predicted by fallback
models, which is not observed \citep{ishi14}.  
A variant   model  includes a   double source   with the CNO elements synthesised by faint SNe and  the   heavier elements made by  conventional core-collapse SNe  \citep{lim03,boni15}.

An alternative, potentially complementary, model for the progenitors of CEMP-no stars is  massive low-metallicity,  fast-rotating stars in the range  40 to 120 M$_{\odot}$, the so-called spinstars, which may experience  efficient mixing and significant mass loss
\citep{Meynet2006A,chiappini2008A&A...479L...9C,maeder2015A&A...576A..56M,Limongi2018}.
Mixing is  driven by rapid rotation, and the diffusion of matter between the H- and He-burning zones largely increases the abundance of  $^{13}$C.  
%In these models, significant  amounts of C and O produced by He-burning in the core are
%transported by mixing processes, mainly shear diffusion,  in the H--burning shell. 
If    $^{13}$C flows into the convective core, it is transported to the surface  and delivered into the interstellar medium by winds \citep{Meynet2006A,chiappini2008A&A...479L...9C,maeder2015A&A...576A..56M,Limongi2018}. 
%Here  the CNO cycles produce 
 %$^{13}$C and $^{14}$N  with amounts that depend on whether the burning is complete or incomplete. 
 %To note that full mixing would   destroy  the $^{13}$C  isotope. 
In the spinstar model, the mixing occurs progressively during the evolutionary stages and $^{13}$C  is lost through  stellar winds, while in the  faint SN model the mixing that produces   $^{13}$C should occur in the explosion and is unconstrained.

 High $^{12}$C/$^{13}$C values   imply   partial or very mild mixing that   occurred either within the star itself or in the progenitor. For  the  unevolved stars    HE\,1327$-$2326  and HE\,0233$-$0343,  all the    $^{13}$C   is coming from  the parent cloud.  Generalising this result  to the giants, it follows  that the  parent generation  of the most metal-poor stars, likely the first stars,    synthesised   significant  amounts of primary $^{13}$C.
A transition seems to occur at [Fe/H] $\approx$ -4  and suggests the presence of   different  progenitors for the two groups of stars. The first generation of massive stars that polluted the gas    seems to have made  less $^{13}$C than   later generations.
  
    Primordial faint SN  models do not make specific predictions concerning the production of $^{13}$C,  while  significant  quantities of $^{13}$C  are expected  from massive, low-metallicity, fast-rotating stars with $^{12}$C/$^{13}$C  ratios between 30 and 300 \citep{chiappini2008A&A...479L...9C}. 
 %   In such a case 
 %faint SNe    could have come first   while {\it spinstars} could have  made   the $^{13}$C observed in UMP stars. 
 %One intriguing possibility is that  the MMP-HMP  stars formed out of a   mono-progenitor  unable to %synthesize  $^{13}$C,  while     UMP are formed from gas    polluted by  multi-progenitors   among which  %some are able to  make significant  $^{13}$C.
 In  the  zero-metallicity non-rotating stars models by \citet{limongi2012ApJS..199...38L},  the  production of primary $^{13}$C, and all CNO isotopes more generally, occurs in a restricted mass range  between 25 and 35 M$\sun,$  where   the convective shell of He, in which $^{12}$C is produced from the He burning,  incorporates the  H shell. Once the He shell begins to engulf H, it extends considerably. The maximum extension of the shell is for a star of  25 M$_{\odot}$;    for more massive stars it decreases  and therefore the abundance of $^{13}$C also decreases. Stars of 20 M$_{\odot}$  or  50 M$_{\odot}$  do not have this mixing  and there is no production of $^{13}$C.  This picture  could  explain the observed trend  with the more massive stars being the parents of the HMP stars and  progressively lower-mass stars progenitors of the  UMP stars. However, the presence of significant rotation could drive extra mixing in layers that are  otherwise in radiative equilibrium. A rotational-driven mixing could bring the $^{12}$C synthesised in the He convective core up into the tail of the H-burning shell, where it is converted into $^{13}$C. However, the production is not monotonic due to a complex interplay  between rotation and convection (Limongi 2023, private communication).
  We note that similar processes necessary to produce $^{13}$C  are also invoked to explain the primary N behaviour at very low metallicities in damped Ly$\alpha$ galaxies \citep{molaro2003astro.ph..1407M,molaro2004oee..sympE..39M,zafar2014MNRAS.444..744Z}.

\begin{figure}
\begin{center}
{\includegraphics[width=60 mm, angle=90,trim={ .cm .cm .3cm 0cm},clip]{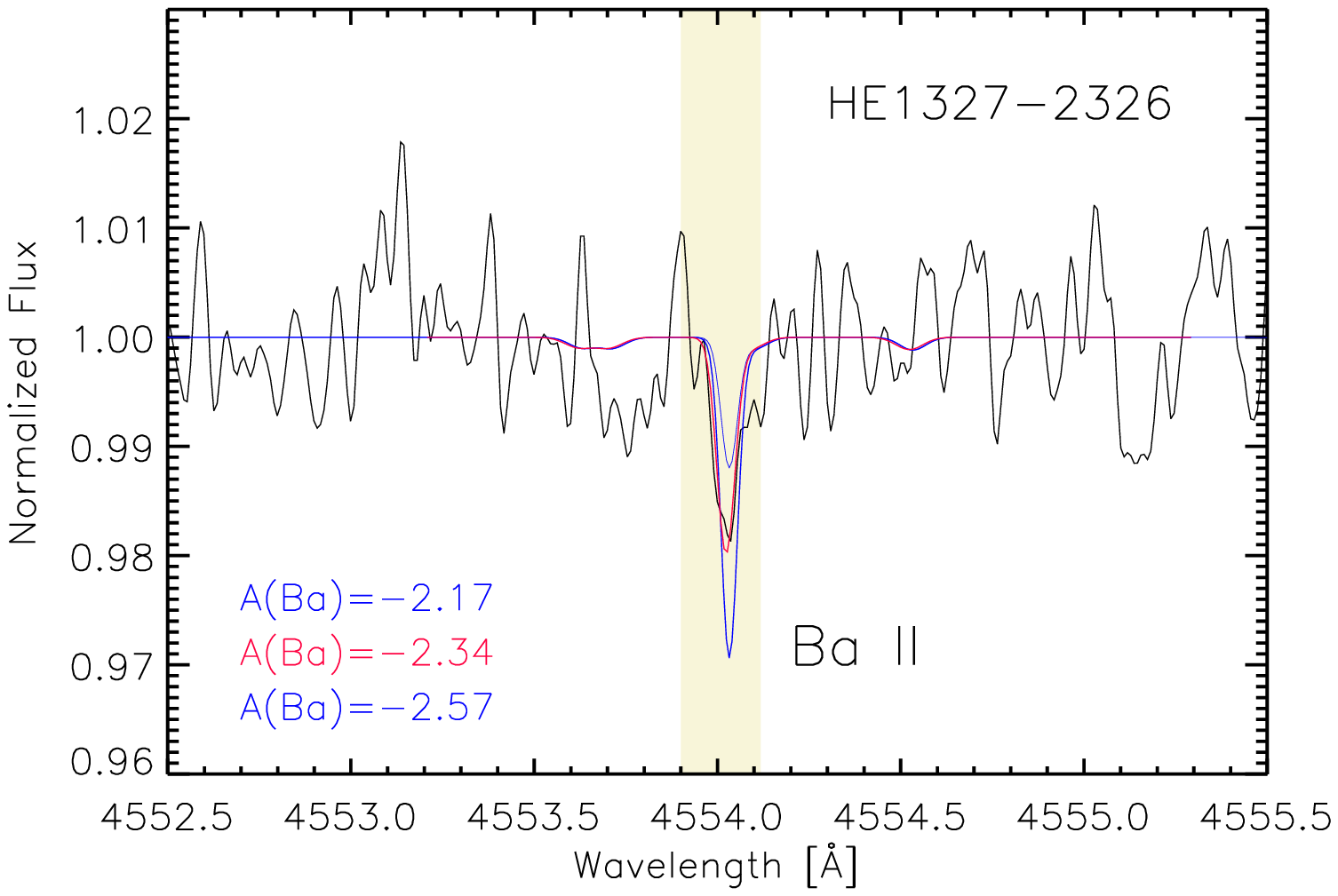}}
\end{center}
\caption{  ESPRESSO spectrum  of HE\,1327-2326  around the \ion{Ba}{II} 455.40 nm line. Three different synthetic models computed with {\tt SYNTHE} are also shown. }
\label{fig:fig4}
\end{figure}

\subsection{n-capture elements}
The CEMP-no classification does not imply a complete absence
of n-capture elements, only that [Ba/Fe]$<$ 0 \citep{chris04}.  For the most iron-poor stars (i.e. [Fe/H] $<$ -5), this condition is difficult to fulfil and limits are much less stringent. There are several cases  where neither Ba nor Fe can be measured, leaving the [Ba/Fe] unconstrained.  For instance, for  SMSS\,J0313-6708 we can place [Ba/H] $<$ -5.6 and [Sr/H] $<$ -5.8,  but the [Ba,Sr/Fe] ratios remain  unbound. For these stars, the  criterion for CEMP-no   membership is necessarily  restricted  to the absolute carbon abundance, which for CEMP-no stars is around A(C) $\approx$ 6.8 $\pm$ 0.5 \citep{boni15}. For  SMSS\,J1605$-$1443, which has  [Fe/H] = -6.21,     \cite{nordlander2019MNRAS.488L.109N} in their discovery paper derived  [Sr/Fe] $<$ 0.2 and [Ba/Fe]$<$1.0. With our ESPRESSO data 
we   slightly improve upon these limits, placing  an upper bound  of [Sr/Fe] $<$ 0.18 and [Ba/Fe]$<$0.6 at the 3$\sigma$ CL. In  HE\,1327-2326 and  HE\,0233--0343,      strontium  has
been measured and found to be relatively abundant, with [Sr/Fe]= 1.04 and 0.3, respectively \citep{fre05,han14}.
The  portion of the ESPRESSO spectrum of HE\,1327-2326 corrected for  a radial velocity of  63.681 \kms  around the \ion{Ba}{II}  455.40 nm  line is shown in Fig. \ref{fig:fig4}; it shows  a hint of a line at the expected position. The abundance would be  [Ba/Fe]=1.3.  Strictly speaking, this   abundance violates  the condition of [Ba/Fe] $<$ 0.0. However,  the  star  shows an   absolute carbon abundance of A(C) = 6.96,   similar to the other CEMP-no stars.  Indeed, there is a small group of  CEMP-no stars with measurable Sr, and some of  them  also have barium measured. With the present detection,  HE\,1327-2326
would be the CEMP-no star with the highest barium abundance relative to  iron. The existence of this group of stars shows that  n-capture elements sometimes need
to be synthesised  at a solar ratio  relative to iron from the very beginning. 
%Sr is an  s-element of the first peak corresponding to  the  magic number of neutrons N=50 while  Ba is  the second peak corresponding to the magic number of N=82. 
This may favour the scenario that foresees the additional contribution from core-collapse SNe in
addition to faint SNe, though the latter are required  to provide the seed nuclei.
For high rotation velocities,  the mixing processes with  successive back and forth motions between the He- and H-burning layers  may also lead to the $^{22}$Ne($\alpha$,n)$^{25}$Mg reaction that produces s-elements by n-captures on seed heavy elements. In this way,  a substantial number of s-elements  of the first peak, such as Sr,  corresponding to  the  magic number of neutrons N=50, are formed, as are   a few  elements of  the second peak, such as Ba, corresponding to the magic number of N=82; nothing of the third peak is formed
 \citep{frishknecht2016MNRAS.456.1803F}.

\begin{figure}
\begin{center}
{\includegraphics[width=85 mm, angle=0,trim={ .cm .cm .cm 0cm},clip]{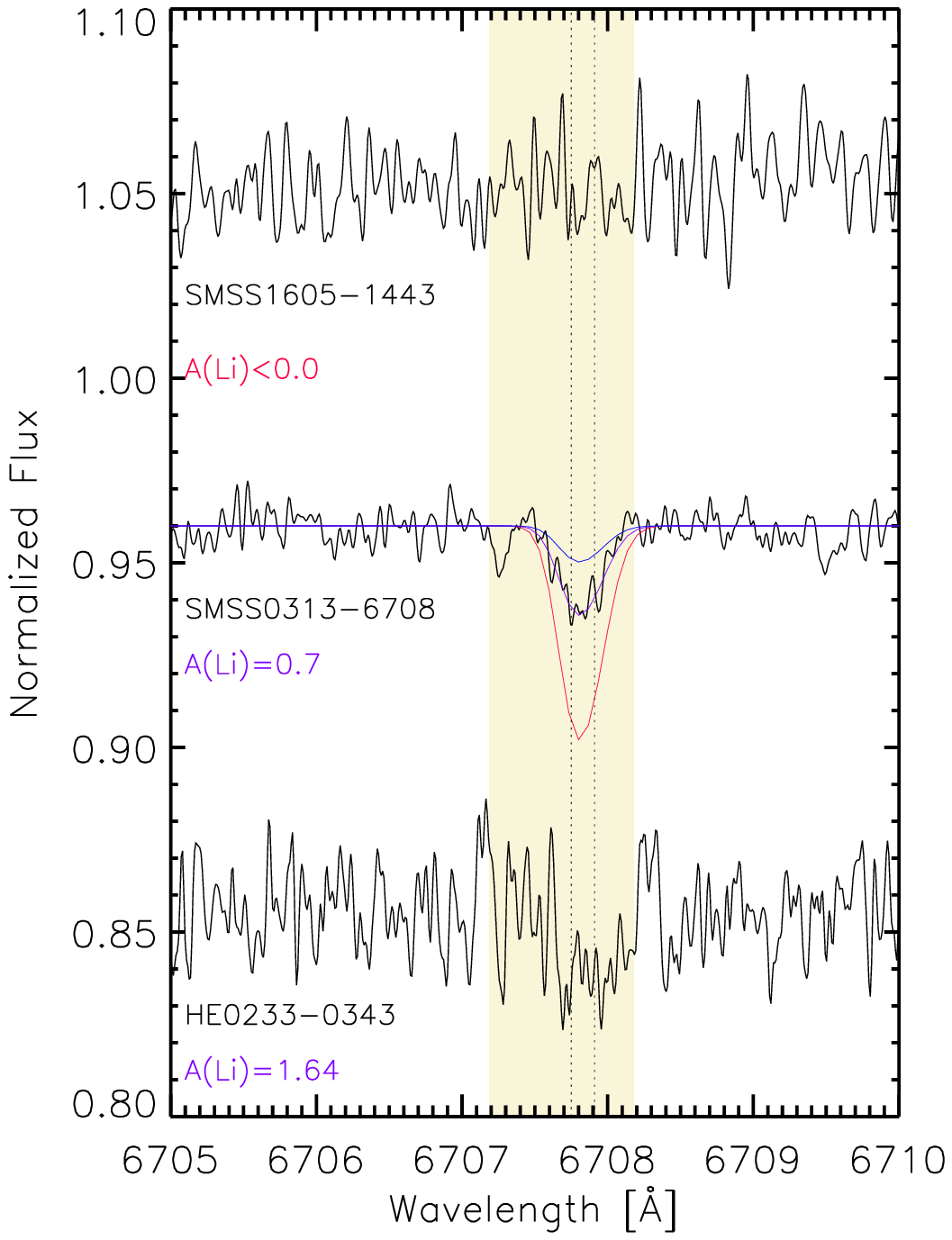}}
\end{center}
\caption{  ESPRESSO spectra for three representative stars around the Li 670.7 nm  position. Radial velocities given in \citet{aguado22} have been subtracted.}
\label{fig:litio_spetra}
\end{figure}

\subsection{Lithium in the CEMP-no stars}

 In the  sample of  stars with [Fe/H]$<$ - 4.5,  there are six unevolved stars with effective
temperatures between 5800 and 6345 K.\ 
As noted by \citet{sbo10}, they  should lie  on the Spite and Spite plateau
at metallicities below [Fe/H] =-3. The HMP and UMP stars follow this trend and magnify
the meltdown,  showing   more Li-depleted stars at the lowest-metallicity end.
Below [Fe/H] $<$ -4.0 there are no stars with Li abundances strictly on the plateau, and about half of them show no detectable Li at all. This is in  contrast with the recent suggestion by \citet{heger2020MmSAI..91...58H}, of significant Li production by neutrino-induced spallation during the explosion of very compact zero-metal core-collapse SNe. The
$^{12}$C/$^{13}$C values could also help in interpreting the origin of Li depletion. 
A clear correlation between $^{12}$C/$^{13}$C and Li abundance is observed in a large sample of disk giants,  with the low  values in the isotopic ratio showing the highest  Li depletion  \citep{takeda2019PASJ...71..119T}. Stars with evidence of mixing  also show Li depletion since Li is burned in the mixing process.  This seems to also hold for the CEMP-no stars. One example is the CEMP-no dwarf CS 22958-042 with $^{12}$C/$^{13}$C = 9 and A(Li) $<$ 0.6 \citep{sivarani2006}.
However, HE\,1327$-$2326  has  A(Li) $<$ 0.62 according to \citet{fre05}, which is considerably below the plateau value of A(Li) $\approx$ 2.2, but has a relatively high value of  $^{12}$C/$^{13}$C. This suggests that the mechanism behind the Li depletion is not    mixing    in the progenitor. \citet{tadafumi2017PASJ...69...24M} also measured Li at the Spite plateau level in two CEMP-no stars with metallicities of about -3.0, arguing that the nature of neither the CEMP-no stars nor their progenitors can explain the higher fraction of Li depletion at low metallicities. The mechanism seems more related to the metallicity, rather than the nature, of the CEMP-no stars or their progenitors.

A similar argument can also be made for the giants. 
 \citet{mucciareelli2022A&A...661A.153M} find a thin Li plateau at A(Li) = 1.09 $\pm$ 0.01 in the giants following the Li dilution, which occurs when the star leaves the main sequence.  It is quite remarkable that in  SMSS\,J0313$-$6708, the most iron-poor star presently known, we measure  A(Li)=0.70 $\pm$ 0.15, in agreement with the 0.7 quoted by \citet{kell12}. After correcting this value using the dilution factor from \citet{mucciareelli2022A&A...661A.153M}, the original Li becomes 0.39 dex  below the primordial value,  regardless of whether this is at the Planck value of A(Li) $\approx$ 2.7 or at the halo stars' value of A(Li) $\approx$ 2.2. Thus,
Li values below  Mucciarelli's plateau indicate the presence of effective Li depletion.  
In the case of SMSS\,J1605$-$1443, we could place 3$\sigma$ upper limits at A(Li) $< $ 0.0, improving upon  the    A(Li) $<$ 0.48 limit from \citet{nordlander2019MNRAS.488L.109N}.  
 In  the three  giants with high  $^{12}$C/$^{13}$C,   Li is observed at A(Li) $<$ 0.6; this is  below    the Mucciarelli plateau, indicating Li depletion in excess of dilution. 
 
\citet{fu2015MNRAS.452.3256F}
suggest a stellar fix to the cosmological Li
problem that involves a substantial pre-main-sequence
depletion partially compensated for  by gas accretion 
with a primordial Li composition of
A(Li)=2.7, as inferred from standard Big Bang nucleosynthesis,  once the baryon density from the cosmic microwave background power
spectrum or the D/H extragalactic measurements is adopted.
 The presence of overshooting,
which is required by helioseismology,
leads to a substantial Li burning in the
pre-main-sequence evolution,   which needs to
be compensated  for with a later accretion  \citep{fu2015MNRAS.452.3256F}.
An accretion rate of 10$^{-8} M_0$ yr$^{-1}$ at
the birth line that then decays exponentially
until it is halted by  photo-evaporation provides
a  self-regulating mechanism able to reproduce
a Li  plateau for a  range
of stellar masses. The MMP and HMP  stars are  smaller and hotter,
and the tenuous accretion disk could be dissipated
before the restoration  of the initial Li is
completed. Thus, a   break of this self-regulating
mechanism could explain, at least qualitatively,
the increased  scatter of the lithium abundances at very low metallicities.

\begin{figure}
\begin{center}
{\includegraphics[width=60 mm, angle=90,trim={ .cm .cm .cm 0cm},clip]{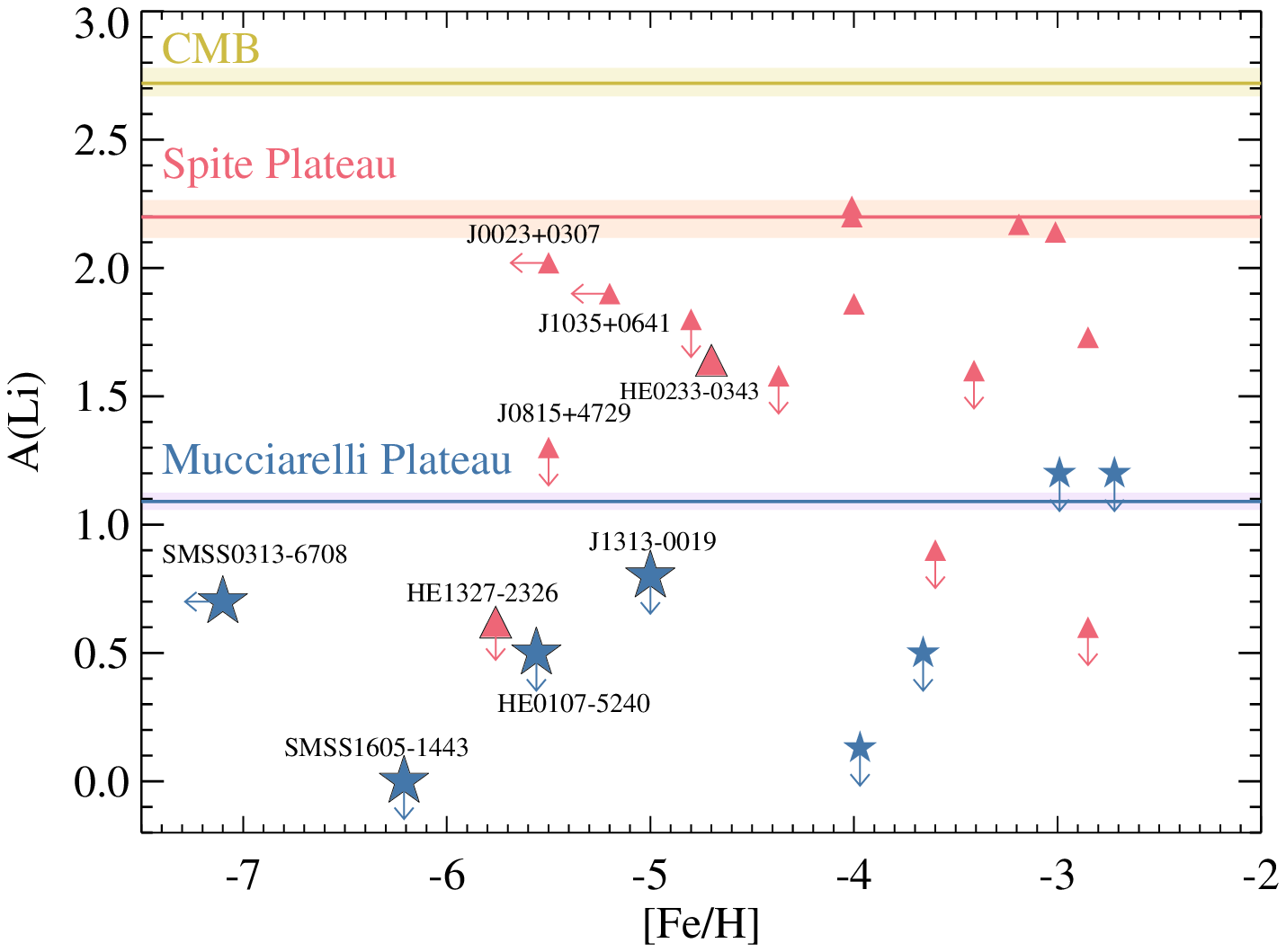}}
\end{center}
\caption{  Lithium abundances for CEMP-no stars from Table \ref{table2}. The two plateaux for dwarfs and giants observed at higher metallicities are shown  in red and blue, respectively. The former is  from \citet{sbo10}, and the latter is at A(Li)=1.09 $\pm$ 0.01 and is from \citet{mucciareelli2022A&A...661A.153M}.  Red triangles are for warm dwarfs, T$_{\rm eff} >$ 5800 K and $\logg > 3.4$. Blue stars are for giants,  $\logg < 3.0$. Supergiants with $\logg < 2.0$ are not shown.}
\label{fig:litio}
\end{figure}

\begin{table*}[t!]  %tbl
\vspace*{0mm}
 \caption{Abundance data for the CEMP-no stars grouped in bins of iron metallicity. Mega metal-poor  (MMP) stars with [Fe/H] $<$ -6.0, hyper metal-poor (HMP) stars with [Fe/H] $<$ -5.0,  ultra  metal-poor (UMP) stars with [Fe/H] $<$ -4.0,  extremely  metal-poor (EMP) stars with [Fe/H] $<$ -3.0, and metal-poor  (MP) stars with [Fe/H] $<$ -1.5.
 %including the two stars with  normal carbon. 
 The  table lists  the CEMP-no stars presently known for which either the carbon isotopic ratio or the Li has been measured in the literature. In boldface are the new measurements presented in this paper. Limits are at the 3$\sigma$ CL. } 
 \label{table2}
\begin{center}\scriptsize
\begin{tabular}{lccccccccl}
Star  &  $ T_{\mathrm{eff}}$ & $ \log g $ &  [Fe/H] &  A(C) & A(Li) & $^{12}$C/$^{13}$C  &     [Sr/Fe] & [Ba/Fe]    & source\\
&   &  &   &  & \\
\hline
MMP&   &  & &   & \\
SMSS\,J0313-6708 &   5125 & 2.3  &  <-7.1 & 6.06& {\bf 0.7$\pm$0.15  } & {\bf 98.5$\pm$5.5} &      &  
& 1,4\\
SMSS\, J1605-1443&4850   & 2.0 &-6.21 & 6.81 & {\bf <0.0 } & {\bf 67.3$\pm$1.0 }&  {\bf <0.18} & {\bf <0.6}&1, 19 \\
&   &  &    &  & \\
\hline
HMP &   &  & & &   &  & \\
HE\,1327-2326 & 6180 & 3.7  & -5.76  &  6.96 & <0.62 & {\bf  > 47 }  &   +1.04&{\bf 1.3 $\pm$0.2 }& 1,  2, 12\\
HE\,0107-5240  &  5100 & 2.2 & -5.56 & 6.75 & <0.50 &   87.0$\pm$6  &    < -0.76    &<0.2  &  2, 5, 12\\
SDSS\,J0023+0307&6140   & 4.8 & <-5.5 & 6.20 & +2.02 $\pm0.08$ && && 17, 24 \\
SDSS\,J0815+4729&6050& 4.6 &-5.50& 7.43 & <1.3& - & <1.02 & <1.91& 18, 23 \\
SDSS\,J1035+0641&6262& 4.0 &<-5.2& 7.02  & +1.90& - &  &&13\\
SDSS\,J1313-0019&5200&2.6&-5.0& 6.42& <0.8& {\bf 39.1$\pm$0.6} &  <-0.28&<0.22&1, 15, 20\\
&   &  & &  & \\
\hline
UMP&   &  & &   & \\
%J1029+1729&5811&4.0&-4.99&<1.0&-&<0.93&<-0.2&-&&&20\\
SDSS\,J0929+0238&5894& 4.5 &-4.97& 7.44 & -& - & <0.70&<1.46& 25 \\
HE\,0557-4840 & 4900 & 2.2  & -4.81  &  5.35   & - &  -  &    <0.0   & < -1.0  & 12\\
 SDSS\,J1742+2531&6345& 4.0 &-4.80&7.26 &  <1.8& - & <0.72 &<1.56&14\\
 HE\,0233--0343&6230&4.43&-4.70& 7.24& {\bf +1.65$\pm$0.2} & {\bf 78.9$\pm$ 5.4}  &   +0.3 &<0.8&1, 7\\
SDSS\,J1442-0015&5850& 4.0 &-4.37& 6.02& <1.58& - &  &&13\\
 HE\,1310--0536&5000&1.9&-4.20& 6.62 & <0.8 & 3&  -1.08$\pm0.14$&-0.50$\pm0.15$& 7\\
 HE\,0057-5959  &  5257 & 2.65 & -4.08  & 5.24 &- &>2 &      &-0.46 & 12\\ 
  SDSS\,J0140+2344 & 5703 & 4.68 & -4.05  &  5.76& +1.86 & >2  &      +1.09& 
  <0.34& 12,13\\
G77-61  &   4000 & 5.05  &  -4.03 & 7.01    & <1.0 & 5$\pm1$  &  & <1.0 & 2, 3\\
HE\,2139-5432 & 5416 & 3.04  & -4.02  & 7.03    & - & >15 &     & 
   <-0.33& 12\\
SDSS\,J1034+0701&6224& 4.0 &-4.01& 6.28& +2.24& - &  &&13\\
SDSS\,J1247-0341&6224& 4.0 &-4.01&  6.64& +2.2:& - &&&13\\
%Pristine 221& 5792& 3.5& -4.66& 1.70&-&<1.87&&&&&21\\
HE\,0134--1519&5500&3.2&-4.00&  5.46  & -&$>$4&  -0.30$\pm0.19$&<-0.50& 7\\
&   &  & &  \\
\hline
EMP &   &  & &  \\
 CS\,22949-037 & 4958 & 1.84  & -3.97  & 5.55    & <0.13  & 4$\pm1$  &   +0.55  &  -0.52& 3,12\\
 HE\,1201-1512 & 5725 & 4.67  & -3.89  & 5.94   & - & >20  &    & <-0.34
  & 12\\
  
HE\,1300+0157& 5529 & 3.25  & -3.75  &  6.02    & +1.0$\pm0.09$ & >3  &   &< -0.85 & 12, 2\\

HE\,2331--7155&4900&1.5&-3.70& 6.10& -&5& -0.85$\pm0.20$&-0.90$\pm0.21$& 7\\
BD\,+44 493 & 5510 & 3.70  & -3.68  & 6.09  & +0.64 & -  &    & -0.60 & 12\\
CS\,22885-096  &   5050 & 1.9  &  -3.66 & 5.40   & <0.5 & - &   & -1.64 & 2\\
SDSS\,J1349+1407&6112& 4.0 &-3.60& 6.82& <0.9& - &  &&13\\
HE\,1506-0113 & 5016 & 2.01  & -3.54  & 6.39   &  +1.01$\pm0.09$ &  >20  &    & -0.80
  & 12\\
Segue 1-7 & 4960 & 1.90  & -3.52  & 7.24   &  - &>50 &    & 
 & 12\\
CS\,29498-043 & 4639 & 1.00  & -3.49  & 6.87   & - &  6$\pm2$ &     & 
 & 2, 12\\
HE\,0146-1548 & 4636 & 0.99  & -3.46  & 6.08 & -& 4  &          & -0.71 & 12\\
%HE1012-1540 & 5745 & 3.45  & -3.47  &- & -  & 2.22 &  1.25 & 2.25  &   &  -0.28& 12\\
HE\,1150-0428 & 5208 & 2.54  & -3.47  & 7.36 &  - & 4  &       &-0.54$\pm0.14$  & 2,12\\
%CS22878-027 & 6319 & 4.41  & -2.51 & -   &- & 0.86    &  <1.06 &    - &  &   & 12\\
%SDSS\,J1507+0051&6555& 4.0 &-3.41&  &  <1.6& &+0.25&&13\\
CS\,30322-023 &   4100 &0.3  &  -3.39 & 5.87  & - & 4 &   & & 2,3\\
BS\,16929-005 & 5229 & 2.61  & -3.34  & 6.11     &  - & >7  &   & -0.41  & 12\\
%CS22166-016 &  5250 & 2.0 &   -2.40& - & -  & 1.02 & - &  - &  &    & 2\\
%HE0007-1832 &   6515 & 3.8  &  -2.72 & - & - & 2.45 & 1.67 & -  & &   & 3\\
CS\,22957-027& 5170 & 2.45  & -3.19  & 7.54   & - & 8$\pm2$  &    &  -1.23$\pm0.21$ & 9,10,8\\
%HE1300-0641  &   5308 & 2.96  &  -3.14 & -  & - & 1.29 & -   & - & & -0.82 & 2\\
%HE1300-2201  &   6332 & 4.64 &  -2.61 & - & - & 1.01 & -  & - & 0.28 &   -0.04& 2\\
   LAMOST\,J1410-0555&6169&4.21&-3.19& 6.80& +2.17&& -0.10&-0.33&22\\
%HE1330-0354  &  6257& 4.13  &  -2.29 & - & - & 1.05 & - & -   & 0.01 &  -0.47& 2\\
HE\,1419-1324  &   4900 & 1.80  &  -3.05 &  7.17 & - & 12$\pm2$ &   & +0.88$\pm0.1$  & 2,3\\
SDSS\,J1424+5615&6088&4.34&-3.01& 6.94& +2.14&-& -0.40&<-0.69&22\\ 
HE\,0100--1622&5400&3.0&-2.90& 8.31& $<$1.12&13&  +0.25$\pm0.25$ &$<$-1.80& 7\\
%HE2142-5656 & 4939 & 1.85  & -2.87  & - & -  & 0.95 &   0.54 & -      &  & -0.63 & 12\\
%HE2202-4831 & 5331 & 2.95  & -2.78 &- & -  & 2.41      & -   & - &  & & 12\\    

&   &  & & &   &  & \\
\hline
MP&   &  & & &   &  & \\
CS\,29502-092 & 5074 & 2.21  & -2.99  & 6.43 &<1.2 & 20  &        & -1.36$\pm0.07$ & 12, 2\\
HE\,2247-7400 & 4929 & 1.56 & -2.87  & 6.29 & - & - &   & -0.94
  & 12\\
CS\,22958-042  & 6250 & 3.5 & -2.85 & 8.76    &  <0.6 & 9$\pm2$ &  & -0.53$pm0.16$
& 2,3, 11\\

CS\,30314-067  &  4400 & 0.7  &  -2.85 & 6.11  & <0.6 & - &   & -0.25  & 2\\
CS\,31080-095  &   6050 & 4.5  &  -2.85 &  8.30   & +1.73 $\pm 0.05$ & >40 &   && 2,3,11\\
CS\,22877-001  &  5100 &   2.2 &  -2.72 & 6.74   & <1.2 &  >10 &    & -0.51& 2,3,8\\
HE\,2319-5228&4900&1.6&-2.60& 7.56& -&5& &&7\\
CS\,22945-017 &   6400 & 3.80  &  -2.52 & 8.22   & - & 6$\pm3$ &   
& 0.55$\pm0.2$&2,3\\
HE\,1410+0213  &   4890 & 2.00  &  -2.52 &  8.27   &  - & 3$\pm 0.5$ &    & +0.05$\pm0.20$ &3\\
CS\,22956-028  &   6700 & 3.50  &  -2.33 &  7.97   & -& 5$\pm2$ &  & +0.16$\pm0.2$  & 3\\
CS\,22887-048&6500&3.7&-1.75& 8.17& -&3& &&8\\ 
   &   &  &    &  & \\
   \hline
\end{tabular}
\end{center}
\vspace*{0mm}
{Ref: 1. this paper; 2. Allen et al. (2012);  3. \citet{masseron2010A&A...509A..93M}; 4. \citet{kel14}, 5 \citet{aguado22}, 6 \cite{Aguado2023A&A...669L...4A}, 7 \citet{hansen2015ApJ...807..173H}}, 8 \citet{aoki2002ApJ...567.1166A}, 9 \citet{bonifacio1998A&A...332..672B}, 10 \citet{norris1997ApJ...489L.169N}, 11 \citet{sivarani2006}, 12 \citet{norris13I}, 13 \citet{boni18}, 14 \citet{boni15}, 15 \citet{alle15}, 16 \citet{caff12I}, 17 \citet{aguado2019ApJ...874L..21A}, 18 \citet{agu18I}, 19 \citet{nordlander2019MNRAS.488L.109N}, 20 \citet{fre15}, 21 \citet{starkenburg14}, 22 \citet{tadafumi2017PASJ...69...24M}, 23 \citet{gonzalez2020sea..confE.142G}, 24 \citet{Frebel2019ApJ...871..146F}, 25 \citet{caffau2016A&A...595L...6C}.
\vspace*{0mm}
\end{table*}

\section{Conclusions}

  By means of  high-resolution spectra   acquired with the ESPRESSO spectrograph at the VLT,  we succeeded in measuring  the $^{12}$C/$^{13}$C   ratio in four of  the lowest-metallicity stars known:  the   MMP   giant  SMSS\,J0313$-$6708, with [Fe/H] $<$ -7.1;   the HMP dwarf  HE\,1327$-$2326, [Fe/H] = -5.8; the giant SDSS\,J1313$-$0019, [Fe/H] = -5.0; and the UMP subgiant  HE\,0233$-$0343, [Fe/H] = -4.7.  We also revised the value for the  star  SMSS\,J1605$-$1443, [Fe/H] = -6.2, turning a lower limit into a value.  Our main results are the following:
 
 \begin{itemize}

 \item{For the three cool    giants SMSS\,J0313$-$6708, SMSS\,J1605$-$1443, and   SDSS\,J1313$-$0019, we derive  a $^{12}$C/$^{13}$C    measurement, while for the two warm unevolved  stars   we provide  a measurement for   HE\,0233$-$0343  and a  lower limit for the  dwarf  HE\,1327$-$2326. Measurements or limits are  all in the range  39 $<$  $^{12}$C/$^{13}$C$<$ 100, with a monotonic decrease going from the more metal-poor stars to the less metal-poor ones.     The values derived for the $^{12}$C/$^{13}$C  ratio provide evidence   of  mixing between the He- and H-burning layers   in the progenitors and a primary production of $^{13}$C at the dawn of  chemical evolution.}
    
   \item{
    For  the  CEMP-no dwarf stars with  $^{12}$C/$^{13}$C values  in the literature, we find that very   low isotopic values, even close to  the CNO cycle   equilibrium value,  are common. In particular, they are also found in  a few dwarfs.  Both the literature $^{12}$C/$^{13}$C values and those obtained here show  a monotonic decrease,   reaching the lowest values  at metallicities in the range  [Fe/H]$\approx$  -4, -3. This   could mark a real  difference between the progenitor $^3$C pollution  captured by   stars with different metallicities. In particular, the decrease  in the $^{12}$C/$^{13}$C values with metallicity could reflect an increase in $^{13}$C  production   by less massive stars \citep{limongi2012ApJS..199...38L}.}
    
    \item{The  ESPRESSO spectrum of HE\,1327-2326    shows a hint of a \ion{Ba}{II} line. If present, the abundance would be  [Ba/Fe] = 1.3, making   HE\,1327-2326
 the CEMP-no star with the highest barium abundance.   Such an    abundance violates  the [Ba/Fe] $<$ 0.0 criterion since HE\,1327-2326  also has a  carbon abundance characteristic of the CEMP-no stars. Indeed, there is a small group  of CEMP-no stars with measured Sr values,  of which   four also have Ba measurements.  The existence of this group of stars within the CEMP-no category indicates a very early synthesis of   n-capture elements.}
    
  \item{  A  general correlation between $^{12}$C/$^{13}$C values and Li abundance holds.  Since the fragile Li  is burned in the mixing process, low   carbon isotopic ratios are generally associated with   the largest  Li depletions \citep{takeda2019PASJ...71..119T}.  This is not observed in our stars, and Li depletion is observed for relatively high  $^{12}$C/$^{13}$C values, suggesting that neither mixing in the progenitor nor the CEMP-no nature of the stars is responsible for the Li depletion. }

\end{itemize}

The study of the isotope ratio in low-mass halo stars has revealed the presence of significant synthesis of $^{13}$C  in the most massive stars of the very first stellar generations that contaminated the gas from which they formed. A synthesis that can only occur through the mixing of the helium and hydrogen burning layers, thus providing important information on the structure of the first stars. Future observations will have to ascertain how the values observed in low metallicity stars can be traced back to the high values observed in stars with solar abundances.

\begin{acknowledgements}

PM acknowledge important discussions with Marco Limongi about the $^{13}C$ production by massive zero metal stars.
DA also acknowledges financial support from the Spanish Ministry of Science and Innovation (MICINN) under the 2021 Ram\'on y Cajal program MICINN RYC2021-032609.
EC acknowledges support from the French National Research Agency (ANR) funded project {\it Pristine} (ANR-18-CE31-0017).
JIGH, CAP, ASM and RR acknowledge financial support from the Spanish Ministry of Science and Innovation (MICINN) project PID2020-117493GB-I00.
MRZO acknowledges financial support from the Spanish Ministry of Science and Innovation through project PID2019-109522GB-C51.
ASM acknowledges financial support from the Spanish Ministry of Science and Innovation (MICINN) under 2018 Juan de la Cierva program IJC2018-035229-I. ASM acknowledge financial support from the Government of the Canary Islands project ProID2020010129.
% This work was supported by FCT - Fundação para a Ciência e a Tecnologia through national funds and by FEDER through COMPETE2020 - Programa Operacional Competitividade e Internacionalização by these grants: UID/FIS/04434/2019; UIDB/04434/2020; UIDP/04434/2020; PTDC/FIS-AST/32113/2017 \& POCI-01-0145-FEDER-032113. 
%This work has made use of data from the European Space Agency (ESA) mission {\it Gaia} (\url{https://www.cosmos.esa.int/gaia}), processed by the {\it Gaia} Data Processing and Analysis Consortium (DPAC, \url{https://www.cosmos.esa.int/web/gaia/dpac/consortium}). Funding for the DPAC has been provided by national institutions, in particular the institutions participating in the {\it Gaia} Multilateral Agreement.
%This work was financed by FEDER–Fundo Europeu de Desenvolvimento Regional funds through the COMPETE 2020–Operational Programme for Competitiveness and Internationalisation (POCI), and by Portuguese funds through
%FCT - Fundação para a Ciência e a Tecnologia under projects POCI-01-0145-
%FEDER-028987, PTDC/FIS-AST/28987/2017, PTDC/FIS-AST/0054/2021 and
%EXPL/FIS-AST/1368/2021, as well as UIDB/04434/2020 \& UIDP/04434/2020,
%CERN/FIS-PAR/0037/2019, PTDC/FIS-OUT/29048/2017.
FPE, CLO, and TMS would like to acknowledge the Swiss National Science Foundation (SNSF) for supporting research with ESPRESSO through the SNSF grants nr. 140649, 152721, 166227, 184618, and 193689. The ESPRESSO Instrument Project was partially funded through SNSF’s FLARE Programme for large infrastructures.
DM is also supported by the INFN PD51 INDARK grant.
This work was financed by FCT - Fundação para a Ciência e a Tecnologia under projects UIDB/04434/2020 \& UIDP/04434/2020, CERN/FIS-PAR/0037/2019, PTDC/FIS-AST/0054/2021, PTDC 2022.04048(Phi in the Sky).
CJM also acknowledges FCT and POCH/FSE (EC) support through Investigador FCT Contract 2021.01214.CEECIND/CP1658/CT0001. MTM acknowledges the support of the Australian Research Council through Future Fellowship grant FT180100194.

\end{acknowledgements}

\bibliography{biblio}

\end{document}